\documentstyle[11pt,aaspp4,rotate]{article}
\def\asec{$^{\prime\prime}$}
\def\and{$\&$ }
\def\kms{km s$^{-1}$}

\begin{document}

\title{NGC 3576 and NGC 3603: Two Luminous Southern HII Regions Observed at High
Resolution with the Australia Telescope Compact Array}
\author{C. G. De Pree$^{1}$, Melissa C. Nysewander$^{1,2}$, W. M. Goss$^{2}$}
\author{$^{1}$Department of Physics \& Astronomy, Agnes Scott College, 141 E. College Ave, Decatur, GA 30030}
\author{$^{2}$National Radio Astronomy Observatory, P.O.Box 0, Socorro, NM 87801}

\begin{abstract}

NGC 3576 (G291.28-0.71; $l$ = 291.3$^{o}$, $b$ = -0.7$^{o}$) and 
NGC 3603 (G291.58-0.43; $l$ = 291.6$^{o}$, $b$ = -0.5$^{o}$) are optically 
visible, luminous HII regions located at distances of 3.0 kpc and 
6.1 kpc, respectively.  We
present 3.4 cm Australian Telescope Compact Array (ATCA) observations of these
two sources 
in the continuum and the
H90$\alpha$, He90$\alpha$, C90$\alpha$ and H113$\beta$
recombination lines with an angular resolution of 7$^{\prime\prime}$
and a velocity resolution of 2.6 km s$^{-1}$.  
All four recombination lines are detected in the integrated profiles of
the two sources.  Broad radio recombination
lines are detected in both NGC 3576 ($\Delta$$V_{FWHM}$ $\geq$
50 km s$^{-1}$) and NGC 3603 
($\Delta$$V_{FWHM}$ $\geq$ 70 km s$^{-1}$). In NGC 3576 a
prominent N-S velocity gradient
($\sim$30 km s$^{-1}$ pc$^{-1}$) is observed,
and a clear temperature 
gradient (6000 K to 8000 K) is found from east to west,
consistent with a known IR color gradient in the source.
In NGC 3603, the
H90$\alpha$, He90$\alpha$ and the H113$\beta$ lines are
detected from 13 individual sources.  The Y$^{+}$ (He/H) ratios
in the two sources range from 0.08$\pm$0.04 to 0.26$\pm$0.10.  
The H113$\beta$/H90$\alpha$ ratio in NGC 3576 is close to the
theoretical value, suggesting that local thermodynamic equilibrium (LTE)
exists.  This ratio is enhanced for most regions
in NGC 3603; enhanced $\beta$/$\alpha$ ratios in other sources
have been attributed to high optical depth or stimulated emission.
We compare the morphology and kinematics of the ionized gas at 3.4
cm with the distribution of stars, 10$\micron$ emission and H$_2$O, OH, and
CH$_3$OH maser emission.  These comparisons suggest that
both NGC 3576 and NGC 3603 have
undergone sequential star formation.
\end{abstract}

\section{Introduction}
NGC 3576 (D = 3.0 $\pm$ 0.3 kpc) and NGC 3603 (D = 6.1 $\pm$ 0.6
kpc) are two of the
highest luminosity optically visible HII regions in our Galaxy 
(Goss \& Radhakrishnan, 1969). Indeed NGC 3603 (with a bolometric 
luminosity of 10$^7$ L$_{\odot}$) contains
some 10$^4$ M$_{\odot}$ of ionized gas, and aside from W49A, may be the most
massive HII region in the Galaxy (Eisenhauer et al. 1998). 
Sources such as these are ideal laboratories to study the formation and evolution
of massive stars and the initial mass function (IMF) of 
giant star forming regions. Similar regions that are visible with the Very Large
Array (VLA) have been studied in detail.
In particular, radio recombination line (RRL) studies of the
Sgr B2 and W49A star forming regions with the VLA (Gaume et al. 1995;
De Pree et al. 1995:
Mehringer et al. 1995) have revealed small-scale outflows, very broad recombination
lines ($\Delta$$V_{FWHM}\sim$80 \kms),
expanding
shells of ionized gas, and extremely compact structures. NGC 3576 and
NGC 3603 appear to be similar regions in terms of luminosity, total mass
and gas kinematics.
Existing optical studies of NGC 3603, for example, have revealed
broad emission lines and a stellar wind bubble, indicative of the
early stages of a star forming region coming into equilibrium with its
environment (Clayton 1986).

Though close in projection ($\sim$40\arcmin), the two
regions are separated by about 30 \kms~in radial velocity as indicated by HI absorption
(Goss \and Radhakrishnan 1969) and hydrogen recombination line 
measurements (Wilson et al. 1970).
Both regions are partially obscured by foreground dust, and studies
of the kinematics of the associated ionized gas have been limited to
optical wavelengths (Clayton 1986; Clayton 1990; Balick et
 al. 1980).  
Thus, existing
velocity fields of the ionized gas are incomplete.  The low resolution
(4\arcmin) 6 cm image of Goss \and Shaver (1970)
shows that the radio emission
associated with the two regions
extends well beyond the detected H$\alpha$ emission.  In both sources,
however, it appears that the large scale ionized structures may be driven by energetic
processes
occuring near their cores.
Evidence from a variety of wavelengths (including the
radio line and continuum data from this study) indicate that both regions
may have undergone sequential star formation, in NGC 3576 from east to west,
and in NGC 3603 from north to south. 

NGC 3603 and NGC 3576 have not been previously imaged with the Australia
Telescope Compact Array (ATCA).
Radio observations have been limited to
low resolution line and
continuum observations such as the $\sim$1$^{\prime}$~
observations of Retallack $\&$
Goss (1980) at 21 cm.
Thus, our 
radio frequency observations ($\theta_{FWHM}\sim$7\arcsec)
improve by factors of $\sim$10 the best spatial resolution
radio observations of the
ionized gas associated with NGC 3576 and NGC 3603.
For both NGC 3576 and NGC 3603, we present high resolution
radio continuum images and radio recombination lines.
New kinematic distances which are in reasonable agreement with the
previously derived spectroscopic distances are derived from the observed
line velocities and a revised Galactic radius of $R_0$=8.5 kpc. We examine the
ionized 
gas morphology and kinematics in the two sources, and possible
star formation histories of these two regions. In addition,
we derive electron temperatures, H113$\beta$/H90$\alpha$ ratios,
helium abundances, and (where possible) carbon line strengths.

\section{Observations and Data Reduction}

Radio continuum and recombination line observations
were made for NGC 3576 and
NGC 3603 at the Australian Telescope Compact Array (ATCA) in the (750 m) B 
and (750 m) D
configuration in two observing runs on the 13th and
19th of June in 1995 with a total of 24 hours of observing time. 
The observing frequency was 8.873 GHz, close to the
the frequency
of the H90$\alpha$ recombination line.  The wide bandwidth
(16 MHz) allowed the simultaneous observation of three additional recombination
lines: He90$\alpha$, C90$\alpha$, and H113$\beta$. 
Detailed observing parameters are presented in Table 1.
The initial calibrations were made using the AIPS
(Astronomical Image Processing
System) software distributed by NRAO.  Each data set was 
flagged 
and the flagged continuum data were then calibrated using the
observed phase and flux density calibrators.  These flags and
calibrations were copied to the line
data, where a bandpass calibration was applied 
and the continuum was subtracted from the $u$-$v$ plane
using the AIPS task UVLIN.  
Continuum images (with angular resolution of $\sim$7\arcsec)
were made
from the line free channels for each source using the AIPS task IMAGR.
This spatial resolution corresponds to a linear size of approximately 
0.09 pc for NGC 3576
($D$= 3.0 kpc) and approximately 0.21 pc for the more distant 
NGC 3603 ($D$= 6.1 kpc).
The noise in the continuum images is dominated by the dynamic
range limitations; thus the rms noise is significantly reduced in the 
continuum-subtracted line data.

After the initial data reduction in AIPS, the line and
continuum data were analysed using 
GIPSY (the Groningen Image Processing System).  Integrated
line profiles were made 
for each pixel above a continuum cutoff
level for both NGC 3576 (5$\sigma$ =
90 mJy beam$^{-1}$) and 
NGC 3603 (5$\sigma$ = 55 mJy
beam$^{-1}$) using the GIPSY task PROFIL.  Gaussian functions 
were then fitted to
the spectral line data.  The output from these fits are the line to
continuum ratio ($T_L/T_C$), the integrated line strength, the
central velocity ($V_{LSR}$)
and the velocity full width at half maximum ($\Delta$$V_{FWHM}$) of
the line emission in each region.
The HII region NGC 3576 consists of a single extended source 
(containing several local bright spots) with
an angular size of $\sim$75\arcsec~($\sim$0.7 pc).
In NGC 3603, thirteen individual sources were observed
above a 5$\sigma$ level in the continuum image.

\section{Results}
\subsection{3.4 cm Continuum Emission}

Continuum images at 3.4 cm were made for both NGC 3576 and
NGC 3603 using the AIPS task IMAGR.
Figure 1 is the resulting 3.4 cm continuum image for NGC 3576.
The crosses in the plot indicate the
positions of the 10 $\micron$ sources tabulated by
Frogel \& Persson (1974).
Table 2 lists the continuum parameters for 
NGC 3576.
These parameters include peak flux density ($S_{peak}$),
integrated flux density ($S_{tot}$), angular diameter in
arcseconds ($\theta$), and linear radius in pc
($r$).
Using the Parkes 64-m telescope, McGee \& Newton (1981) report a
14.7 GHz flux density of 97.8 Jy in NGC 3576. 
We detect about 73\% of the single dish flux density, or
71$\pm$7 Jy. From the 13 sources in NGC 3603, we detect a
total flux density of 25 Jy, or 38\% of the reported single
dish flux density of 65.9 Jy.

Figure 2 shows the 3.4 cm continuum image of NGC 3603.
The crosses indicate the positions of the
known 10 $\micron$ sources in NGC 3603 (Frogel et al. 1977).
Each emission peak higher than the continuum 5$\sigma$ cutoff
(55 mJy beam$^{-1}$) has been designated as a separate region.
The regions are labeled A to M, corresponding to increasing right
ascension.
The AIPS task JMFIT fitted two
dimensional Gaussians to the individual continuum
sources in NGC 3603.
Table 2 presents continuum parameters for each region
(A-M) in NGC 3603. 
These parameters include peak flux density ($S_{peak}$),
integrated flux density ($S_{tot}$), deconvolved geometrical mean
diameter in arcseconds
($\theta$), and the linear radius in pc
($r$).
Table 3 presents derived physical parameters for the regions in
NGC 3603. These
calculated parameters include electron density ($n_e$), emission measure
($EM$), mass of ionized gas ($M_{HII}$), excitation parameter ($U$), 
Lyman continuum photon flux ($N_{LyC}$) and continuum
optical depth ($\tau_c$). 
The continuum parameters were calculated assuming a uniform
density, spherical, ionization bounded HII region, using the corrected
formulae of Mezger \& Henderson (1969). The calculated values are of
course dependent on these assumptions, and in a complex region
like NGC 3603, these assumptions may be overly simplistic.
Figures 1 and 2 are plotted in
B1950 coordinates.  The last two figures in the paper (Figures 8 and
9) are plotted in J2000 coordinates
to help in comparisons between the radio
continuum emission and images at other frequencies, since images
in the literature are shown in both epochs.
 
\subsection{Radio Recombination Lines}
\subsubsection{NGC 3576}
A continuum-weighted, integrated line profile was generated for NGC 3576 
using the GIPSY task PROFIL. The integrated profile
was generated for all pixels above a continuum 5$\sigma$ cutoff level. 
In addition, integrated profiles were generated for each pixel above the 
continuum 5$\sigma$ level in a 7\arcsec
$\times$7\arcsec~ box centered on each of the positions of the
Irs sources tabulated by Frogel \& Persson (1974).
The integrated source profile, and the four profiles
line profiles were then fitted with Gaussian functions using the
GIPSY task PROFIT.
Figure 3a presents the integrated line profile and model fit for
NGC 3576.  Each of the four lines (H90$\alpha$,
He90$\alpha$, C90$\alpha$ and H113$\beta$) is clearly detected, and
labeled in Figure 3a.
The integrated spectra for the Irs sources (Irs1/2, Irs3, Irs4 and Irs5;
Frogel \& Persson 1974) in NGC 3576
are shown in Figures 4a-d.  
The spectra in Fig. 3a and Figs. 4a-d show
the data (points), Gaussian fits (solid
line), and the
residuals (dotted line) for each recombination line. 
The parameters of Gaussian fits to the lines 
(V$_{LSR}$, $\Delta$$V_{FWHM}$, and
T$_l$/T$_c$) are presented in Table 4. 

From the
integrated emission from NGC 3576, we determine a helium abundance
of $Y^+$=0.09$\pm$0.01, and a H113$\beta$/H90$\alpha$ ratio close to the
theoretically predicted value of 0.276 ($\beta$/$\alpha$=0.25$\pm$0.01).
The electron temperature in NGC 3576 (T$_e^*$) is calculated 
assuming local thermodynamic equilibrium
(LTE) from both the H90$\alpha$ and
H113$\beta$ recombination lines (using equation [22] from Roelfsema \&
Goss [1992]):
\begin{equation}
T_e^* = (6943\nu^{1.1}{{T_c}\over{T_l\Delta V_{FWHM}}}{{1}\over{1+Y^+}})^{0.87} K,
\end{equation}
with line width ($\Delta V_{FWHM}$) given in \kms and $\nu$ in GHz.
These LTE temperatures are given in Table 4.
An electron temperature ($T_e$) may also be
calculated with corrections made for non-LTE effects (using Eq. [23]
from Roelfsema \& Goss [1992]):
\begin{equation}
T_e=T_e^*[b_n(1-{{\beta_n\tau_c}\over{2}})]^{0.87} K,
\end{equation}
where $b_n$ and $\beta_n$ values are from the tables of
Salem \& Brocklehurst (1979).

For NGC 3576, a turbulent velocity ($v_{turb}$) and an electron temperature
($T_e$)
are calculated by solving two equations for 
line width (Eq. 3) using the
hydrogen and helium (H90$\alpha$ and He90$\alpha$) recombination line widths. 
The line width can be expressed as
\begin{equation}
<v_{turb}^2>={{3\Delta~V_{H}^2}\over{8ln2}}-{{3kT_e}\over{M}}
\end{equation}
where the $v_{turb}$ is the turbulent velocity in \kms, $k$ is the
Boltzmann constant, $M$ is the mass of the atom, $T_e$ is the
electron temperature, and $\Delta$$V_{H}$ is the line FWHM
in \kms.
Using this method for
NGC 3576, we derive a turbulent velocity $v_{turb}$=17$\pm$1.0 \kms~
and an electron temperature of (T$_e$) of 6100$\pm$200 K.

In order to examine the spatial variation in the H90$\alpha$
line parameters, single
Gaussian fits were made to each pixel above the continuum 5$\sigma$ level in NGC 3576.
The results of these point-to-point fits to the H90$\alpha$ line are presented
in Figures 5a and 5b.  Figure 5a shows the spatial
variation in the the central line  
velocity ($V_{LSR}$), with continuum contours overlaid on the
greyscale representation of the ionized gas velocity.
Figure 5b shows the spatial variation in the linewidth
($\Delta$$V_{FWHM}$) of the H90$\alpha$ line. 
For NGC 3576, a map of the LTE electron temperature was also generated,
using Equation 1 (above) for each point above 10$\sigma$
in the continuum image.
The resulting LTE temperature image is shown in Figure 5c, indicating
a clear E-W temperature gradient.
This temperature gradient has also been plotted using the AIPS task
IRING, which averages emission in concentric semicircles.
The semicircles (with a thickness of $\sim$7\arcsec), 
are centered on the continuum
peak of NGC 3576 and extend to the east.
A plot of temperature as a function of distance from the peak (in arcsec) is
presented as Figure 6. The apparent temperature gradient is discussed
in \S 4.2.1.

\subsubsection{NGC 3603}
An integrated line profile was generated for 
the NGC 3603 region above
a 5$\sigma$ cutoff in the continuum emission.
Figure 3b
presents the integrated line profile and model fit for
NGC 3603. Because of the large variation in line velocities
across the source, the carbon line (apparent in some of the individual
source profiles in NGC 3603) was fitted with the line width fixed at
10 \kms.
Gaussian fits were made to the integrated
profile, and
the parameters of these fits
are given in Table 5.
Figures 7a-7m present the radio recombination line
profiles and Gaussian fits for each of the thirteen 
subregions (A-M) in NGC 3603. For all 13 subregions,
the signal-to-noise ratio is sufficiently high to
fit the H90$\alpha$, He90$\alpha$, and H113$\beta$ lines.
The C90$\alpha$ recombination line was successfully fit only in regions
NGC 3603 B and H with a fixed carbon line width. 
Sources in which parameters were fixed and the
fixed values are
indicated parenthetically in Table 5.
In a few cases, the fitted 
linewidth ($\Delta$$V_{FWHM}$) of the He90$\alpha$ line may be broad due
to blending with the C90$\alpha$ line in the sources where a C90$\alpha$
line was not separately fitted. 
The line parameters from the Gaussian fits to the integrated emission from 
NGC 3603 and its 13 subregions 
are presented in Table 5. As with NGC 3576, the table also includes
the LTE electron temperature ($T_e^*$), the corrected electron temperature
($T_e$), the ionized helium to hydrogen ratio ($Y^+$), and the H113$\beta$ to H90$\alpha$ ratio ($\beta/\alpha$), and the turbulent velocity ($v_{turb}$). 
For NGC 3603, the non-LTE electron temperature was calculated
using Equation 2 (above).
The turbulent velocity was calculated from the width of the
H90$\alpha$ line and the non-LTE electron temperature using Eq. 3 (above).
These temperatures and turbulent velocities are given in Table 5.


\section{Discussion}

\subsection{Distances to NGC 3576 and NGC 3603}

McGee \& Newton (1981)
detected recombination lines at 14.7 GHz from
25 high emission measure HII regions using the Parkes
64-m radio telescope. Both NGC 3576 (1109-610)
and NGC 3603 (1112-610) were included in their survey. McGee \&
Newton detected the H76$\alpha$ line from NGC 3576 and NGC 3603
at -24.1$\pm$0.1 \kms~ and 7.7$\pm$0.1 \kms~ respectively. In addition
to H76$\alpha$, McGee \& Newton detected the He76$\alpha$ line toward
both sources, and the carbon line toward NGC 3576. 
For NGC 3576, the detected line width in the H90$\alpha$ 
line ($\Delta$$V_{FWHM}$=28.7$\pm$
0.1) is in good agreement with the H76$\alpha$ line width
published by McGee \& Newton ($\Delta$$V_{FWHM}$=29.1$\pm$0.1). 
The integrated H90$\alpha$ line
width for NGC 3603 ($\Delta$$V_{FWHM}$=36.9$\pm$0.25) is 
in reasonable agreement with
the value for the H76$\alpha$ line given by McGee \& Newton ($\Delta$$V_{FWHM}$=41.9$\pm$0.2). 

The distance to NGC 3576 has been discussed at length in 
two papers, Goss \& Radhakrishnan (1969), and Persi et al (1994).
Goss \& Radhakrishnan found a kinematic distance of 3.6 kpc,
just beyond the tangential point of the Galaxy,
using the Schmidt galactic model and a value of R$_{0}$ = 10 kpc. 
This distance is derived from a line 
center velocity of -23 km s$^{-1}$, and circular motion within
the galaxy.  This kinematic distance agrees well with the
findings from Caswell \& Haynes (1987). 
Our new observations, using a revised $R$$_{0}$ = 8.5 kpc, and a
line velocity of -20.0 $\pm$ 0.1 km s$^{-1}$
suggest a distance of 3.0$\pm$0.3 
kpc for NGC 3576.
Because no ionizing cluster is optically visible, no reliable
spectroscopic distance has been determined.  
Humphreys (1972), in a study of the effect of streaming in
the Carina arm, determined a distance of 3.2 kpc for NGC 3576, and detected 
both velocity variations ($\sim$4 km s$^{-1}$ )
and a large spread in the individual velocities of early type stars
in this direction.
Persi et al. (1994) derive a distance based on the star HD 97499
(position plotted in Figure 8).
However, the high resolution radio continuum image of the source (Fig. 8)
shows that there may not be a direct association between HD 97499 and NGC 3576.
HD 97499 (RA = 11 12 10.1, 
Dec = -61 18 45.1 J2000) is located $\sim$2\arcmin~east of the
peak in the 
3.4 cm continuum emission.  As a result, it is unlikely that HD 97499
contributes greatly to the ionization seen in NGC 3576.
Finally, the stellar velocity of HD 97499
(-32 km s$^{-1}$ $\pm$ 11 km s$^{-1}$; Crampton, 1972)
does not agree well with the observed velocities in the ionized gas
(-23 km s$^{-1}$
[McGee \& Gardner, 1968] and -25 km s$^{-1}$ [Caswell \& Haynes,
1987]). Thus, distances associated with HD 97499 should not simply
be applied to NGC 3576.

The determination of the distance to NGC 3603 is aided by the presence 
of an optically visible ionizing cluster.  Distance estimates made from
stellar observations can be compared to kinematic distances derived
from the ionized gas. The earliest spectroscopic
distance estimate was made by Sher (1965) at 3.5 kpc.  Without a large
number of faint stars visible, the spectroscopic 
distance estimate was admittedly
crude.  Shortly thereafter, Goss \&
Radhakrishnan (1969) found a distance of 8.4 kpc based on the observed
line velocity of 10 km s$^{-1}$, using 
R$_{0}$ = 10 kpc.  Caswell \& Haynes (1987) also found a
kinematic distance using R$_{0}$ = 10.5 kpc and $V$ = +11 km s$^{-1}$
of 8.6 kpc.
McGee \& Newton (1982) report an H76$\alpha$
line velocity of 7.7$\pm$0.1 \kms.
Our velocity for the integrated
line in NGC 3603 is $V_{LSR}$=9.1$\pm$0.1, consistent
with previously reported values. 
The H90$\alpha$ velocities found for the thirteen individual
regions range between $-$2 km s$^{-1}$ and 19 km s$^{-1}$. 
Using
the velocity from the integrated line ($V_{LSR}$=9.1 km s$^{-1}$) and
R$_{0}$ = 8.5 kpc, we determine a kinematic distance of 6.1$\pm$0.6 kpc.

{\it UBV} stellar photometry of NGC 3603
has resulted in distance estimates as low as 3.5 kpc
(Sher, 1965) and as high as 8.1 kpc
(Moffat, 1974).
Melnick \& Grosb$\phi$l (1982) derive a distance of 5.3 kpc.
Moffat (1974), using the stars observed by Sher (1965) and
a different reddening scale 
derive a distance of 8.1 $\pm$ 0.8 kpc.
Finally, Melnick, Tapia \& Terlevich (1989) use {\it UBV} photometry of
the ionizing cluster of NGC 3603 to find a distance modulus
of ($m-M)_0$=14.3, which corresponds to a distance of $\sim$7.2 kpc.
Their value is in good agreement with both Moffat (1974), and the
kinematic distances of Goss \&
Radhakrishnan (1969), and 
Caswell \& Haynes (1987), and in reasonable agreement with the
distance derived from our integrated radio recombination line
velocity.
The smaller distances inferred from the measurements of
Sher (1965) and Melnick \& Grosb$\phi$l (1982) are likely the result of
errors in photometry due to nebular contamination
(Melnick, Tapia \& Terlevich 1989).

\subsection{NGC 3576}
NGC 3576 
has been extensively observed in the
radio continuum (Goss \and Shaver 1970) and in radio
recombination lines (McGee \and Gardner 1968, McGee \& Newton 1981, Wilson
et al. 1970).  Recombination line studies of the region show the line velocities to be
centered at approximately $-$23 \kms.
The core (diameter$\sim$1.5\arcmin) of NGC 3576
is the location of
five identified infrared (10$\micron$) sources 
(Frogel \and Persson 1974), as well
as CH$_3$OH (Caswell et al. 1995)
and H$_2$O (Caswell et al. 1989) masers.  The presence of bright 10$\micron$
emission, water masers and compact thermal radio
emission are typical indications
of the early stages of
star formation and dense circumstellar environments.

Persi et al. (1994) have identified 135 individual stars with 
IR excess (to distinguish cluster stars from red field stars)
in an extremely
young galactic cluster associated with the HII region NGC
3576, using $JHK$ images and photometry.
Approximately 40 of these IR sources are 
members of the cluster.
In their $JHK$ images, 
Persi et al. (1994) detect a color gradient in the near-IR,
indicating increasing extinction toward the western edge of
NGC 3576. This gradient suggests
that star formation may have progressed in NGC 3576
from east to west, with the youngest stars currently located at the western
edge of the source.  

In addition, Persi et al. (1994) provide a summary of existing studies of the
region, and an overlay of the 21 cm continuum
(Retallack \& Goss 1980), the 10 $\micron$ continuum sources
(Frogel \& Persson 1974) and their K band mosaic of the region.  The
overlay (their Fig. 7) clearly shows that the radio continuum, near-IR
continuum and 10$\micron$ continuum are all strongly peaked toward  the
western edge of the source. Our 3.4 cm radio continuum image (Fig. 1)
improves
significantly 
on the resolution of Retallack \& Goss (1980), and shows that there
is a coincidence between the location of the infrared sources (Irs1-Irs5) tabulated by
Frogel \& Persson (1974) and the extended ionized radio continuum
emission.  Figure 1 shows the 3.4 cm radio continuum with crosses and
labels indicating the positions of the
10$\micron$ peaks (Frogel et al. 1977).  
All five sources are coincident with the ionized
gas emission and are 
near local peaks in the radio continuum emission.

\subsubsection{Characteristics of the Ionized Gas}
Our radio recombination line observations have allowed us to examine
in detail the variation in electron temperature and helium abundance
in NGC 3576.  Our results in both cases seem to strengthen the case
for sequential star formation in that source.
The LTE electron temperatures show a
clear gradient from east to west, with temperature
increasing with distance (east) from the continuum peak.
The gradient is apparent in Figures 5c and 6.  Figure 5c
shows the spatial variation in electron temperature across the source.
Figure 6 is a plot of temperature as a function of distance from the
3.4 cm continuum peak.
The LTE electron temperature  
increases with distance from 6000 K to a maximum of $\sim$8,000 K, and then falls to
lower levels in the more diffuse gas to the east.
If the ionizing stars are those IR sources coincident with the radio
continuum, this rising temperature signature is expected (Hjellming 1966).
Lower energy photons
on average would be absorbed first, with the higher energy
photons penetrating deeper into the diffuse gas. Thus, the temperature
gradient confirms a scenario in which the youngest, hottest stars in the
HII region are located near the sharp western edge of the radio continuum.

Under LTE conditions, the theoretical value of the H113$\beta$ to H90$\alpha$
hydrogen ratio is 0.276 (Shaver \& Wilson, 1979).  Deviations from this
theoretical ratio indicate local deviations from LTE, and the
ratio can be used to correct the derived electron temperature.
The H113$\beta$ to H90$\alpha$ ratio in NGC 3576
varies from $\sim$0.25$\pm$0.05 to $\sim$0.35$\pm$0.05, 
with the lowest values associated with the bright peak in the
radio continuum along the western edge of the HII region. Pressure
broadening can lower the $\beta$/$\alpha$ ratio from the theoretical
value.
When the $\beta/\alpha$ ratio is lower 
than the theoretical value, electron temperatures derived
from the $\beta$ lines will of course
be higher than those derived from $\alpha$
lines. In fact, for NGC 3576, the electron temperatures derived from
the H113$\beta$ line are $\sim$11 \% higher than those derived
from the H90$\alpha$ recombination line.  

A small number of Galactic objects are known to have helium abundances
in excess of 20\%. Such elevated helium abundances have been reported in the
W3 core (Adler, Wood \& Goss 1996, Roelfsema, Goss \& Mallik
1992), and are thought to be associated with the 
environments of evolved stellar objects that are rapidly losing mass.
A small but significant positive gradient in helium abundance
(Y$^{+}$) is detected from the western peak of the source
(Y$^{+}$ = 0.08 $\pm$ 0.01)
to the more diffuse ionized gas to the east (Y$^{+}$ = 0.14 $\pm$ 0.03).
This range is consistent with the $Y^+$ value reported by McGee \&
Newton (1981) of N(He$^+$)/N(H$^+$)=0.091.
The abundance gradient is consistent with the suggestion of
Persi et al. (1994) who detect a color 
gradient in the near-IR indicating increasing extinction
to the west of the source. Persi et al. (1994) propose that
the youngest stars are located at the western edge of the ionized
gas.
The higher helium
abundances found to the east of NGC 3576 may indicate the
presence of a population
of older stars and their associated mass loss.

\subsubsection{Gas Kinematics in NGC 3576}
Figures 5a and 5b show that both broad lines
and velocity gradients are present in the ionized gas associated with
NGC 3576. 
Double-peaked emission lines are located over
a $\sim$15\arcsec~region at the northern edge of the source.
In Fig. 5a, the double-peaked lines have been fitted
with a single Gaussian with a $\Delta$$V_{FWHM}$
= 50 \kms. Closer inspection (see Figure 1 inset) shows that
locally the lines are possibly double-peaked.
The double-peaked recombination lines in this region have
a velocity separation of $\sim$25
km s$^{-1}$, and may be due to either
the expansion of an ionized shell of gas, or the outflow of ionized
gas from a local source.
The broad line region is located to the east of the Irs1/Irs2 peak
(Frogel \& Persson 1974).
The region of the broadest radio
recombination lines
corresponds spatially to a distinct gap in the 10 $\mu$m emission
between Irs1/Irs2 and Irs5.

If the split lines are due to expansion, then the rate of expansion
is $\sim$13 km s$^{-1}$. Double-peaked lines with similar velocity
separations have been seen in some of the compact HII regions
in Sgr B2 North (De Pree et al. 1996).  The spatial resolution
of the current data is insufficient to determine whether an
expanding shell of ionized gas is located at this position.
Alternatively, the 
double-peaked lines may arise near the base of a large scale
ionized 
outflow.

A large, approximately north-south velocity gradient of 30 km s$^{-1}$
pc$^{-1}$ is observed in the ionized gas, and is shown
in Figure 5a.
The velocity across the source varies from $\sim$0 km s$^{-1}$
along the southern edge to $\sim$-35 km s$^{-1}$ in the north.
The gradient is
not likely due to rotation in the ionized gas, since these rotational
velocities would imply an approximate
enclosed mass of M=2$\times$10$^5$ M$_{\odot}$. The
gradient is  more likely due to an ionized outflow, and may
in fact be the base of a much larger scale outflow.
A wide field of view optical image of the region
(Fig. 8; Persi et al. 1994) shows several large loops and
filaments of ionized gas extending several minutes of arc to the
north of NGC 3576. 

\subsubsection{Sequential Star Formation in NGC 3576}
Figure 8 shows the positions of cluster members
with infrared excess (crosses; Persi et al. 1994)
overlaid on the 3.4 cm radio continuum.
Approximately 40 infrared excess sources have been identified.
Eight of the stellar positions are coincident with the radio
continuum emission, with the majority of the sources
located to the north, east and south of the radio continuum.
Only a single infrared excess source is located to the west 
of the radio continuum peak. This spatial distribution would be
expected if star formation were proceeding from east to west
into the molecular cloud. Water (H$_2$O) and methanol (CH$_3$OH)
maser emission (plotted) are located coincident with the peak in the radio
continuum emission.

The location of molecular gas in NGC 3576 is also consistent
with a triggered star formation scenario. The peak in the 1.0 mm dust
continuum emission (Cheung et al. 1980) is located to the west of
the bright radio continuum peak. The 1.0 mm emission traces the core of the
molecular cloud located to the west of
NGC 3576. Similar distributions of ionized
and molecular gas are found near the Sgr B2 F HII 
regions (De Pree et al. 1996),
G34.3 (Gaume et al. 1994), and others. In fact, the morphology
of NGC 3576 is ``cometary'', and closely resembles a mirror image
of the cometary source G34.3 (Gaume et al. 1994) in which a dense
molecular cloud core is located at the eastern periphery of the
ionized gas.

Taken as a whole, the radio, infrared, optical and molecular
imaging of NGC 3576 suggest that the oldest ionizing sources
are located to the east of the region, and the youngest
ionizing sources to the west.  
The distribution of stars, ionized gas, molecular gas, and 
water masers
in NGC 3576 is reminiscent of simple scenarios
of sequential star formation.  In this picture, the star
formation process slowly ``eats away'' at a dense cloud, so that the
most evolved stars are located far from the cloud core. The ionized gas
is found at the periphery of the cloud, and the youngest sources
(protostars)
are still embedded in the nascent molecular material, but reveal their
presence with water maser emission. In many such regions in the
Galaxy, the stellar population is
not observable.  The ability to observe stars, ionized gas and molecular
gas in this distant source make it unique in the Galaxy.

\subsection{NGC 3603}
NGC 3603 fills a gap in luminosity
between a region like Orion, with $\sim$1 O-type star, and R136 in 30
Doradus, with $\sim$100 O-type stars (Eisenhauer et al. 1998). 
The ionizing core of NGC 3603 (HD 97950) consists of approximately
20 O-type and Wolf-Rayet stars (Moffat, 1983).
From photometric measurements, 
Melnick, Tapia \& Terlevich (1989) have identified
181 sources (50 with $U$, $B$, and $V$ photometry) associated 
with the stellar cluster within
NGC 3603.  The authors found an
average age of 2.5 $\pm$ 2 Myr and evidence that
star formation
propagated from north to south, ending at the bordering
molecular cloud (Cheung 1980).
Brandner et al. (1997) have identified a ring-shaped nebula and bipolar
outflows from Sher 25, a cluster member of HD 97950.  Sher 25 (a B1.5Iab
supergiant) is
located approximately $\sim$20" north of the core of the cluster.
Brandner et al. present evidence for association of this star with HD 97950
and compare the star to the progenitor of SN 1987A.

The initial mass function (IMF) of HD 97950, the stellar
cluster associated with NGC 3603, has been discussed in recent
articles by Moffat et al. (1994), Hofmann
et al. (1995), Zinnecker (1995) and Eisenhauer et al. (1998). 
Because of the proximity 
of NGC 3603, high resolution, deep studies allow detailed
counts of stellar mass distribution to be made. 
Moffat et al. (1994) observed the central cluster with the
HST/PC, studying its dense stellar core and the population of Wolf-Rayet stars. Moffat et al. (1994) conclude that in terms of density and
stellar population, NGC 3603 is a `Galactic clone' of R136 in
30 Doradus.
Hofmann et al. derive an IMF power law of $\Gamma$  
=$-$1.4 $\pm$ 0.6 ($\xi\propto$M$^{\Gamma}$)
for 30-60 $M$$_{\odot}$ which is similar to that of 30 Doradus.
Using speckle masking techniques with the HST, Hofmann et al. (1995)
found a similar slope for stellar masses of 15 to 50 $M$$_{\odot}$
of $\Gamma$ =$-$1.6 $\pm$ 0.2.  Both Eisenhauer et al. and Zinnecker 
use adaptive
optics to extend the IMF to stellar masses less than 1 $M$$_{\odot}$.  
Some
theoretical and observational studies have claimed that starburst
systems are deficient in low mass stars, but Eisenhauer et al. (1998)
find that the IMF follows a Salpeter power law with an index of
$\leq-$0.73, and 
that there does not appear to be
a low mass truncation to the IMF down to the observational limit of 0.2 $M$$_{\odot}$.
From such properties of NGC 3603, it may be possible to 
extrapolate properties of larger extragalactic starbursts.

Sequential star formation has also been suggested in NGC 3603,
proceeding from north to south (Melnick, Tapia \& Terlevich 1989).  The
presence of evolved objects at the north end of the source
and embedded IR sources and a molecular cloud to the south
qualitatively confirms a sequential star formation scenario
(Hofmann et al. 1995).
Investigations of NGC 3603
have proceeded on two fronts: radio and optical observations
of the ionized gas, and optical and infrared observations of the
exciting stars. While the stars and gas in the 
giant HII region are visually observable, these high resolution
radio observations present the first unobscured view of the
ionized gas on small scales.
Melnick, Tapia \& Terlevich
(1989) have tabulated the positions of 181 stars in NGC 3603
from their CCD {\it UBV} photometry. The positions of the 50
cluster member stars for which
$U,~B,$ and $V$ magnitudes are tabulated are shown in Figure 9.  
The figure shows the positions of these stars (crosses) plotted
with the 3.4 cm radio continuum.
Coordinates for the stellar positions
were derived from pixel coordinates using the known position of 
the B1.5Iab supergiant (Sher 25)
as given by Brandner et al. (1997).

Optical studies indicate both the presence of high velocity
gas and of highly evolved stars. Clayton (1986) first investigated the velocity
structure in the ionized gas located at the core of NGC 3603, and
found both large scale motions (perhaps kinematic evidence of
earlier supernovae) and two smaller shells, possibly wind-driven
bubbles. The H$\alpha$ and [NII] line profiles were made along two
lines across the core of NGC 3603. One possible explanation of the
wind-driven bubbles was suggested by Moffat et al. (1994) who
detected three Wolf-Rayet (WR) stars in their HST/PC
observations of the cluster core. Hofmann et al. (1995) confirm the
presence of three, perhaps four WR stars in the cluster core. Brandner
et al. (1997) provide fascinating evidence that one of the sources
believed to be an older member of the cluster (Sher 25, a B1.5Ia supergiant)
is at the center of a clumpy ring of optical emission and a possible
bipolar outflow.  They compare the source to the progenitor
of SN1987A.

Caswell \& Haynes (1987) detected strong
hydrogen radio recombination line emission from NGC 3603.
The source was also included in the recombination
line survey of McGee \& Newton (1981), and both hydrogen and
helium (at 14.7 GHz) were detected. Our integrated H90$\alpha$ line
width ($\Delta$$V_{FWHM}$=36.9$\pm$0.25) is in reasonable agreement with
the value for the H76$\alpha$ line given by McGee \& Newton ($\Delta$$V_{FWHM}$=41.9$\pm$0.2). 

The present study of the ionized gas in NGC 3603 addresses a number of the
outstanding issues in this source: 

1. Are there any regions of enhanced helium abundance,
as might be expected in the presence of 3-4
WR stars?

2. Is there any evidence in the ionized gas emission of the
bipolar outflow associated with the B1.5 supergiant
Sher 25 (Brandner et al. 1997)?

3. How does the location of ionized gas compare to the known positions of
infrared sources (Frogel et al. 1977), WR stars (Moffat et al. 1994 and
references therein) and maser emission (Caswell et al. 1989, Caswell et al. 1998)?

\subsubsection{Helium Abundance and $\beta$/$\alpha$ Ratios}
McGee \& Newton (1981) derived a helium abundance in NGC 3603 of
Y$^{+}$ =0.069 (no uncertainty given). Our calculated helium abundance from the
integrated line profile is significantly higher (Y$^{+}$ =0.13$\pm$0.01),
but in their notes on individual sources, McGee \& Newton (1981) indicate
that their He76$\alpha$ line profile was rather noisy.

Moffat et al. (1994) detect three prominent
Wolf-Rayet stars in NGC 3603 in their narrowband HeII (4686 A)
images.  WR stars have been cited as the possible explanation of
enhanced helium abundances (Adler, Wood \& Goss 1996).  It is certainly reasonable to look for
evidence of such enhanced abundances in the vicinity of known WR
stars. 
In fact, the highest helium abundances in NGC 3603 are detected in
source C, which has a value of Y$^{+}$ =
0.26 $\pm$ 0.10. Interpretation of this high value is complicated
by the large uncertainty.
The ionized gas in NGC 3603 is located on the periphery of the WR wind-blown
bubble, so it is possible that ionized helium ($Y^+$) abundances
are enhanced by the presence of an evolved star. 

The observed H113$\beta$/H90$\alpha$ ratios for the 13 
sources in NGC 3603 range from
0.29 $\pm$ 0.04 to 0.53 $\pm$ 0.12, all above the theoretical
value ($\beta$/$\alpha$=0.276).  Most detections are within
2$\sigma$ of this theoretical value, but a few sources 
(NGC 3603 E, H, I, and K) exceed the
theoretical value by more than 3$\sigma$.
Anomalously high $\beta$/$\alpha$ ratios
were found from H92$\alpha$ and H115$\beta$ recombination line observations in
two Galactic center regions (G0.18-0.04;  the Sickle and the Pistol; Lang, et al., 1997).
Thum et al. (1995) propose a model that can explain enhanced $\beta$/$\alpha$
ratios as the result of high continuum optical depth.  However, NGC 3603
A-M appear to have high $\beta$/$\alpha$ ratios, and yet be low density
($n_e\leq$10$^4$ cm$^{-3}$) regions with $\tau_c\leq$1.  Cersosimo \&
Magnani (1990) suggest that under such conditions, high $\beta$/$\alpha$
ratios may result from stimulated emission caused by a background source.

\subsubsection{Ionized Gas Kinematics}
No ionized gas is detected above a 5$\sigma$ continuum level
at the location of Sher 25.
The optical study of Clayton (1986) indicates clearly that there
are both large scale and small scale motions in the ionized gas. Plate
1 of Clayton shows an H$\alpha$ image of the central 6\arcmin~ of
NGC 3603, and the positions of the two slit positions through the core.
The bright ionized gas observed in our high resolution image
(see Fig. 1b and Fig. 9) lies to the south of the cluster
core. 
On scales probed by our high resolution observations, only 
one source (NGC 3603 F) has clearly double-peaked recombination
line emission.  
Two Gaussians with velocities separated by $\sim$29 \kms (V$_{LSR}$ 
=27.5 \kms~and $-$1.1 \kms) have been fitted
to the H90$\alpha$ line in NGC 3603 F. The velocity
separation is similar to that seen in other
Galactic wind-driven bubbles and also in NGC 3576. 
In the radio continuum, source F is a relatively faint, marginally resolved
source located near the western edge
of the southern bright region. NGC 3603 F has  a deconvolved diameter
of $\theta_{FWHM}\sim$25\arcsec,
giving the source a linear diameter of
0.5 pc (D = 6.1 kpc). Thus, source F may be an
expanding shell of ionized gas. 
A small optically observable wind-driven "stellar bubble" 
with $r\sim$0.6 pc has 
been observed centered on the HD97950
star cluster in 
NGC 3603
(Balick, Boeshaar \& Gull, 1980).

In NGC 3603, the individual source velocities (as
derived from the H90$\alpha$ line)
range from $-$3.7$\pm$0.3 km s$^{-1}$
in source A to to 15.9$\pm$0.2 \kms~in source I. 
The broadest lines are detected 
in NGC 3603 F ($\Delta$$V_{FWHM}$=52.7$\pm$2.2 \kms). As indicated
above, the broad lines in this source likely result from the
double-peaked lines apparent in Fig. 7f.
Broad lines can result from a variety of local conditions
including pressure broadening or spatially unresolved
velocity gradients.  Given the relatively high frequency of
the observations and 
other evidence for velocity gradients in the source, it is likely that
the broad lines in source M (located to the north of the cluster
core) are related to gas kinematics.
NGC 3603 M has $v$$_{turb}$ $\simeq$ 26 $\pm$ 10 km
s$^{-1}$.
Other localized regions in the source
have very broad recombination line emission. Lines as
broad as $\Delta$$V\sim$80 \kms~
are detected in the low surface
brightness gas between sources NGC 3603
D and F.  And in the bright core of NGC 3603 D ($\sim$0.07 pc in
diameter) line widths approach $\Delta$$V\sim$75 km s$^{-1}$. In
a sequential star formation scenario, sources D and F are located
at the boundary of the molecular cloud and the stellar cluster, and 
would be in the early stages of their evolution, perhaps still
coming into equilibrium with their environment or photoevaporating
material from the molecular cloud interface. 

\subsubsection{Evidence for sequential star formation in NGC 3603}
In Figure 2, the positions of ten 10$\micron$ sources (Frogel et al. 1977)
are plotted over the 3.4 cm continuum.
Figure 9 shows the positions of cluster stars and known maser emission.
Error bars for the absolute positions of the IR sources are $\pm$10\arcsec,
and the size of the crosses in Fig. 2 indicates the absolute error in the positions
of the infrared sources. In several cases (e.g. Irs 1, 2, 9, 14), the infrared sources are
coincident with radio continuum sources.

Perhaps the clearest evidence for sequential star
formation in NGC 3603 is the relative location of the radio emission
from the ionized gas, the stellar positions, and the position of
embedded infrared sources and maser emission.  The distribution is
very much like that observed in NGC 3576.  The most evolved
stars are located to the north, centered in what appears at
many frequencies to be the wind-blown bubble resulting from the
WR stars located there. The molecular cloud core is located farthest
to the south, and the ionized gas emission (apparent at optical and
radio frequencies) is located at the periphery between the stars and
the cold gas.

\section{Conclusions}
Both NGC 3576 and NGC 3603 provide rare opportunities to
study massive star formation in a variety of spectral regions. Massive
star forming regions like W49A and Sgr B2 are impressive at radio
frequencies, but the stellar populations are inaccessible in the
infrared.  In these
two sources, high resolution stellar photometry and radio continuum
and recombination line observations provide a more complete picture
of the physical conditions and complex gas kinematics.

The following is a summary of our major conclusions:

(1) In these high resolution radio continuum and recombination line
observations, we have identified a number of previously
unresolved radio continuum sources, particularly in NGC 3603.

(2) The observed temperature gradient in NGC 3576 is in agreement
with the known color gradient in the source and supports a
sequential star formation scenario that has been proposed for the
region.

(3) We have detected several compact sources
in NGC 3603 with broad recombination lines. One source
(NGC 3603 F) has double peaked profiles, with the peaks separated by
$\sim$30 \kms.

(4) The brightest radio continuum emission in NGC 3603 is located to
the south of the core of the stellar cluster HD 97950.  There is no radio
continuum or line emission apparent at the position of Sher 25.

(5) Derived helium abundances and electron temperatures in NGC 3576 are in 
good agreement with values
given in McGee \& Newton (1981)
while those in NGC 3603 are significantly higher.

(6) Enhanced helium abundances ranging in value from $Y^+$=0.16 to 0.26 are 
detected in 
NGC 3603 A-D and J.  
These regions
may have had their helium abundance enriched by the 3 or 4 known WR sources
located at the cluster core.

(7) In both sources, the relative positions of stars, ionized gas,
molecular gas and masers is suggestive of a sequential star formation
scenario.  In NGC 3576 the star formation appears to have proceeded
from east to west, in NGC 3603 from north to south.

\begin{acknowledgements}
C. G. De Pree gratefully acknowledges the support of an American Astronomical
Society Small Research Grant. W. M. Goss would like to thank R. D. Ekers, J.B. Whiteoak, \& D. McConnell for support during a sabbatical visit to the Australia
Telescope Compact Array in 1996. The Australia Telescope is funded by the
Commonwealth of Australia for operation as a National Facility operated by
CSIRO. The National Radio Astronomy Observatory is a facility of the
National Science Foundation, operated under cooperative agreement by Associated
Universisites, Inc. The authors also thank H. Zinnecker, J. Caswell,
W. Brandner, P. Palmer, D. Mehringer, L. Rodriguez, R. Gaume, M.
Claussen and an anonymous referee for helpful discussions and comments.
\end{acknowledgements}


\begin{table}
\centering
\clearpage
\caption{Observational Parameters of NGC 3576 and NGC 3603}
\begin{tabular}{lcc}\\\hline\hline
\\
Parameter & NGC 3576 & NGC 3603 \\
\\
\hline
\\
Date (configuration)......................... & 13 June 1995 (750m D) & 13 June 1995 (750m D) \\
& 19 June 1995 (750m B) & 19 June 1995 (750m B) \\
Total observing time (hr)................. & 12 & 12 \\
R.A. of field center (J2000).............. & 11$^{h}$ 11$^{m}$ 51.88$^{s}$ & 11$^{h}$ 15$^{m}$ 02.6$^{s}$ \\
Decl. of field center (J2000)............. & -61$^{o}$ 18$^{\prime}$ 31$^{\prime\prime}$ & -61$^{o}$ 15$^{\prime}$ 46$^{\prime\prime}$ \\
Major axis ($^{\prime\prime}$).................................. & 7.0\asec & 7.0\asec \\
Minor axis ($^{\prime\prime}$).................................. & 6.8\asec & 6.9\asec \\
Position angle ($^{o}$)............................. & -9$^{o}$ & -13$^{o}$ \\
LSR central velocity (km s$^{-1}$).......... & -23 & 10 \\
Total bandwidth (MHz)................... & 16 & 16 \\
Number of channels.......................... & 256 & 256 \\
Channel Separation (kHz, km s$^{-1}$)... & 62.5 (2.1) & 62.5 (2.1) \\
Spectral resolution (kHz, km s$^{-1}$)... & 75.6 (2.6) & 75.6 (2.6) \\
Continuum noise (mJy beam$^{-1}$)................ & 18 & 11 \\
Line noise (mJy beam$^{-1}$)........................... & 2.1 & 2.2 \\
\\
\hline
\end{tabular}
\end{table}

\clearpage
\begin{table}
\centering
\scriptsize
\caption{Continuum Parameters for NGC 3576 and NGC 3603}
\begin{tabular}{rrrcccc}\\\hline\hline
Source & RA & Dec & $S_{Peak}$ & $S_{Tot.}$ & $\theta$ & $r$ \\
& (J2000) & (J2000) & (Jy beam$^{-1}$) & (Jy) & ($^{\prime\prime}$) & (pc) \\ \hline\hline
NGC 3576 & 11$^h$ 11$^m$ 51.32$^s$ & -61$^o$ 18$\arcmin$ 43\farcs5 & 4.1$\pm$0.4 & 71$\pm$7 & $\sim$90  & 1.4 \\ 
& & & & \\
NGC 3603 A & 11$^h$ 14$^m$ 56.02$^s$ & -61$^o$ 13$\arcmin$ 55\farcs8 & 0.10$\pm$0.01 & 1.3$\pm$0.3 & 24 & 0.52 \\
B & 14$^m$ 56.51$^s$ & 13$\arcmin$ 17\farcs0 & 0.08$\pm$0.01 & 2.5$\pm$0.4 & 38 & 0.80 \\
C & 15$^m$ 00.90$^s$ & 13$\arcmin$ 32\farcs7 & 0.06$\pm$0.01 & 0.85$\pm$0.3 & 24 & 0.52 \\
D & 15$^m$ 01.99$^s$ & 16$\arcmin$ 33\farcs0 & 0.11$\pm$0.01 & 1.2$\pm$0.3 & 22 & 0.46 \\
E & 15$^m$ 02.62$^s$ & 15$\arcmin$ 53\farcs8 & 0.40$\pm$0.02 & 1.9$\pm$0.2 & 13 & 0.28 \\
F & 15$^m$ 06.13$^s$ & 16$\arcmin$ 40\farcs5 & 0.12$\pm$0.01 & 1.7$\pm$0.3 & 25 & 0.52 \\
G & 15$^m$ 08.76$^s$ & 16$\arcmin$ 55\farcs7 & 0.30$\pm$0.01 & 4.4$\pm$0.3 & 26 & 0.54 \\
H & 15$^m$ 09.45$^s$ & 16$\arcmin$ 41\farcs4 & 0.23$\pm$0.01 & 4.6$\pm$0.3 & 30 & 0.64 \\
I & 15$^m$ 10.36$^s$ & 16$\arcmin$ 17\farcs4 & 0.25$\pm$0.02 & 1.3$\pm$0.2 & 14 & 0.30 \\
J & 15$^m$ 14.24$^s$ & 17$\arcmin$ 34\farcs0 & 0.11$\pm$0.01 & 2.2$\pm$0.3 & 30 & 0.65 \\
K & 15$^m$ 18.75$^s$ & 16$\arcmin$ 58\farcs1 & 0.08$\pm$0.01 & 1.2$\pm$0.3 & 25 & 0.53 \\
L & 15$^m$ 24.20$^s$ & 12$\arcmin$ 54\farcs0 & 0.06$\pm$0.01 & 0.94$\pm$0.3 & 29 & 0.61 \\
M & 15$^m$ 31.59$^s$ & 13$\arcmin$ 16\farcs7 & 0.06$\pm$0.01 & 1.1$\pm$0.3 & 26 & 0.55 \\
\hline\hline
\end{tabular}
\end{table}

\clearpage
\begin{table}
\centering
\scriptsize
\caption{Derived Continuum Parameters for NGC 3603}
\begin{tabular}{lcccccc}\\\hline\hline
Source 	& $n_e$ & $EM$	& $M_{HII}$ & $U$ & $N_{LyC}$ & $\tau_{c}$ \\
& (10$^3$ cm$^{-3}$) & (10$^6$ pc cm$^{-6}$) & (M$_{\odot}$) & (pc cm$^{-2}$) & (10$^{48}$ s$^{-1}$) & \\ \hline\hline
A & 0.92 & 0.9 & 13 & 50 & 5.6 & 0.005 \\
B & 0.67 & 0.7 & 35 & 62 & 11 & 0.004 \\
C & 0.75 & 0.6 & 10 & 44 & 3.7 & 0.005 \\
D & 1.0 & 1.0 & 10 & 49 & 5.1 & 0.006 \\
E & 2.8 & 4.6 & 6 & 57 & 8.2 & 0.027 \\
F & 1.0 & 1.1 & 15 & 54 & 7.1 & 0.007 \\
G & 1.6 & 2.8 & 25 & 76 & 19 & 0.016 \\
H & 1.3 & 2.1 & 32 & 76 & 20 & 0.012 \\ 
I & 2.1 & 2.7 & 5 & 50 & 5.4 & 0.016 \\
J & 0.87 & 1.0 & 23 & 60 & 9.5 & 0.006 \\
K & 0.85 & 0.8 & 12 & 48 & 5.0 & 0.005 \\
L & 0.66 & 0.5 & 15 & 47 & 4.8 & 0.003 \\
M & 0.72 & 0.6 & 12 & 45 & 4.0 & 0.003 \\
\hline
\end{tabular}
\end{table}


\clearpage
\begin{table}
\centering
\scriptsize
\caption{Recombination Line Parameters and Derived
Quantities for or NGC 3576}
\begin{tabular}{lcccc}\\\hline\hline
& $V$ & $\Delta$$V$ & $T$$_{l}$/$T$$_{c}$ & T$_{e}$$^{*}$ \\
& (km s$^{-1}$) & (km s$^{-1}$) & & (K) \\
\hline
H90$\alpha$ & -20.0$\pm$0.1 & 28.7$\pm$0.1 & 0.105$\pm$0.001 & 6300$\pm$700 \\
He90$\alpha$ & -19.3$\pm$0.4 & 24.8$\pm$0.8 & 0.011$\pm$0.001 & \\
C90$\alpha$ & -21.5$\pm$0.9 & 8.75$\pm$1.8 & 0.003$\pm$0.001 & \\
H113$\beta$ & -22.3$\pm$0.2 & 33.6$\pm$0.4 & 0.022$\pm$0.001 & 7100$\pm$800 \\
& & & & \\
H90$\alpha$ & & & & \\
{\it Irs1/Irs2} &  -24.0$\pm$0.1 & 29.1$\pm$0.1 & 0.100$\pm$0.001 & 6500$\pm$700 \\
{\it Irs3} &  -17.2$\pm$0.1 & 26.5$\pm$0.1 & 0.112$\pm$0.001 & 6400$\pm$700 \\
{\it Irs4} &  -14.5$\pm$0.1 & 32.0$\pm$0.3 & 0.081$\pm$0.001 & 7200$\pm$800 \\
{\it Irs5} &  -18.8$\pm$0.2 & 36.2$\pm$0.4 & 0.098$\pm$0.002 & 5500$\pm$600 \\
\hline\hline
\end{tabular}
\end{table}


\clearpage
\begin{table}[ht]
\centering
\scriptsize
\caption{Recombination Line Parameters and Derived Quantities for NGC 3603}
\begin{tabular}{llcccccccc}\\\hline\hline
Source & & $V$ & $\Delta$$V$ & $T$$_{l}$/$T$$_{c}$ & T$_{e}$$^{*}$ & T$_{e}$ &
 Y$^{+}$ & $\beta$/$\alpha$ & $V$$_{turb}$ \\
& & (km s$^{-1}$) & (km s$^{-1}$) & & (K) & (K) \\
\hline

Total & H90$\alpha$ & 9.14$\pm$0.1 & 36.9$\pm$0.25 & 0.075$\pm$0.001 & 6600$\pm$500 & & & & 24$\pm$2  \\
& He90$\alpha$ & 8.9$\pm$1.0 & 38.9$\pm$2.6 & 0.009$\pm$0.001 & & & 0.13$\pm$0.01 \\
& C90$\alpha$ & 6.2$\pm$1.9 & 10.0 (fixed) & 0.002$\pm$0.001 & & & & \\
& H113$\beta$ & 5.9$\pm$0.3 & 42.6$\pm$0.8 & 0.026$\pm$0.001 & 4800$\pm$400 & & & 0.39$\pm$0.01 \\
\\

A & H90$\alpha$ & 8.08$\pm$0.2 & 26.3$\pm$0.5 & 0.107$\pm$0.003 & 6300$\pm$1100 & 6700$\pm$1200 & & & 14$\pm$3 \\
& He90$\alpha$ & 4.0$\pm$1.3 & 25.0$\pm$3.1 & 0.019$\pm$0.003 & & & 0.17
$\pm$0.03\\
& H113$\beta$ & 2.5$\pm$0.8 & 31.6$\pm$1.9 & 0.033$\pm$0.003 & 4900$\pm$900 & 5200$\pm$1000 & & 0.31$\pm$0.03 \\
\\

B & H90$\alpha$ & -3.65$\pm$0.3 & 28.3$\pm$0.7 & 0.103$\pm$0.004 & 6300$\pm$1600 & 6400$\pm$1600 & & & 17$\pm$4 \\
& He90$\alpha$ & 0.9$\pm$2.0 & 28.8$\pm$4.8 & 0.016$\pm$0.004 & & & 0.16
$\pm$0.04 \\
& C90$\alpha$ & -12.5$\pm$1.3 & 10 (fixed) & 0.015$\pm$0.003 \\
& H113$\beta$ & -7.4$\pm$1.1 & 29.0$\pm$2.6 & 0.029$\pm$0.004 & 6000$\pm$1500 & 6100$\pm$1500 & & 0.29$\pm$0.04 \\
\\

C & H90$\alpha$ & 3.8$\pm$0.7 & 33.2$\pm$1.7 & 0.104$\pm$0.007 & 4300$\pm$1700 & 4200$\pm$1700 & & & 22$\pm$9 \\
& He90$\alpha$ & 9.1$\pm$4.3 & 44.1$\pm$10 & 0.020$\pm$0.006 & & & 0.26
$\pm$0.10 \\
& H113$\beta$ & -5.2$\pm$4.1 & 49.1$\pm$10 & 0.023$\pm$0.006 & 4400$\pm$1700 & 4300$\pm$1700 & & 0.32$\pm$0.11 \\
\\

D & H90$\alpha$ & -0.80$\pm$0.4 & 35.9$\pm$1.1 & 0.078$\pm$0.003 & 6400$\pm$2000 & 6800$\pm$2100 & & & 23$\pm$7 \\
& He90$\alpha$ & 1.1$\pm$2.2 & 32.8$\pm$5.6 & 0.014$\pm$0.003 & & & 0.16
$\pm$0.05 \\
& H113$\beta$ & -1.7$\pm$1.0 & 31.4$\pm$2.6 & 0.029$\pm$0.003 & 5700$\pm$1800 & 6100$\pm$1900 & & 0.33$\pm$0.05 \\
\\

E & H90$\alpha$ &  1.9$\pm$0.2  &  41.9$\pm$0.5  &  0.066$\pm$0.001  & 6700$\pm$600 & 8500$\pm$800 & & & 27$\pm$3  \\
& He90$\alpha$ &  -2.2$\pm$1.5  &  36.9 (fixed)  &  0.008$\pm$0.001  & & & 0.11$\pm$0.01 \\
& H113$\beta$ &  -2.3$\pm$0.6  &  46.5$\pm$1.4  &  0.022$\pm$0.001  & 5300$\pm$500 & 6700$\pm$600 & & 0.36$\pm$0.02 \\
\\

F & H90$\alpha$ & 11.5$\pm$0.8 & 52.7$\pm$2.2 & 0.056$\pm$0.003 & 6500$\pm$3300 & 7000$\pm$3600 & & & 36$\pm$19 \\
& He90$\alpha$ & 27.6$\pm$2.7 & 22.7$\pm$6.9 & 0.011$\pm$0.005 & & & 0.08$\pm$0.04 \\
& H113$\beta$ & 5.9$\pm$2.4 & 59.9$\pm$6.8 & 0.020$\pm$0.003 & 4700$\pm$2400 & 5100$\pm$2600 & & 0.40$\pm$0.08 \\
\\

G & H90$\alpha$ & 15.5$\pm$0.4 & 25.6$\pm$1.0 & 0.115$\pm$0.006 & 6200$\pm$2200 & 7900$\pm$2800 & & & 13$\pm$5 \\
& He90$\alpha$ & 20.0$\pm$1.6 & 17.2$\pm$3.8 & 0.024$\pm$0.001 & & & 0.14$\pm$0.05 \\
& H113$\beta$ & 8.3$\pm$1.2 & 31.2$\pm$3.1 & 0.040$\pm$0.006 & 4400$\pm$1500 & 5600$\pm$1900 & & 0.43$\pm$0.07 \\
\\

H & H90$\alpha$ & 11.8$\pm$0.1 & 32.2$\pm$0.3 & 0.082$\pm$0.001 & 6900$\pm$1200 & 8300$\pm$1400 & & & 19$\pm$3 \\
& He90$\alpha$ & 12.4$\pm$0.9 & 29.3$\pm$2.2 & 0.011$\pm$0.001 & & & 0.12$\pm$0.02 \\
& C90$\alpha$ & 8.2$\pm$1.6 & 10.0 (fixed) & 0.003$\pm$0.001 \\
& H113$\beta$ & 8.5$\pm$0.4 & 37.9$\pm$0.9 & 0.028$\pm$0.001 & 5100$\pm$900 & 6300$\pm$1100 & & 0.40$\pm$0.02 \\
\\

I & H90$\alpha$ & 15.9$\pm$0.2 & 34.8$\pm$0.6 & 0.068$\pm$0.002 & 7900$\pm$2200 &  9300$\pm$2600  & & & 21$\pm$6 \\
& He90$\alpha$ & 7.9$\pm$3.1 & 44.9$\pm$8.0 & 0.006$\pm$0.001 & & & 0.11$\pm$0.03 \\
& H113$\beta$ & 13.4$\pm$0.7 & 39.6$\pm$1.8 & 0.023$\pm$0.002 & 5900$\pm$1600 &
 6800$\pm$1800 & & 0.39$\pm$0.03 \\
\\

J & H90$\alpha$ & 7.3$\pm$0.3 & 39.9$\pm$0.9 & 0.085$\pm$0.003 & 5700$\pm$1600 & 5800$\pm$1600  & & & 26$\pm$7 \\
& He90$\alpha$ & 2.9$\pm$3.6 & 59.0$\pm$9.7 & 0.011$\pm$0.003 & & & 0.18$\pm$0.05 \\
& H113$\beta$ & 5.3$\pm$1.2 & 47.8$\pm$2.9 & 0.031$\pm$0.003 & 3600$\pm$1000  &
 3700$\pm$1000 & & 0.43$\pm$0.05 \\
\\

K & H90$\alpha$ & 5.5$\pm$0.3 & 40.8$\pm$0.7 & 0.093$\pm$0.002 & 5100$\pm$1000 &  5000$\pm$1000  & & & 28$\pm$6 \\
& He90$\alpha$ & 7.9$\pm$1.9 & 31.2$\pm$4.7 & 0.012$\pm$0.003& & & 0.10$\pm$0.02 \\
& H113$\beta$ & 3.8$\pm$0.7 & 43.8$\pm$1.9 & 0.037$\pm$0.002 & 3600$\pm$700 & 3500$\pm$700  & & 0.43$\pm$0.03 \\
\\

L & H90$\alpha$ & 6.4$\pm$0.6 & 38.2$\pm$1.5 & 0.142$\pm$0.008 & 3800$\pm$1000 & 3700$\pm$1000  & & & 26$\pm$7 \\
& He90$\alpha$ & 5.7$\pm$4.5 & 35.0 (fixed) & 0.018$\pm$0.004 & & & 0.11$\pm$0.03 \\
& H113$\beta$ & 2.7$\pm$1.3 & 32.9$\pm$3.1 & 0.063$\pm$0.008 & 2900$\pm$800 &  2800$\pm$800  & & 0.37$\pm$0.07 \\
\\

M & H90$\alpha$ & 4.6$\pm$0.5 & 37.3$\pm$2.1 & 0.115$\pm$0.009 & 3600$\pm$1300 &  3500$\pm$1300  & & & 26$\pm$10 \\
& He90$\alpha$ & 8.5$\pm$6.3 & 33.8 (fixed) & 0.015$\pm$0.005 & & & 0.11$\pm$0.04 \\
& H113$\beta$ & 4.3$\pm$2.1 & 44.9$\pm$5.4 & 0.051$\pm$0.009 & 2600$\pm$1000 &
2500$\pm$1000 & & 0.53$\pm$0.12 \\
\\
\hline
\end{tabular}
\end{table}
\newpage
\centerline{FIGURE CAPTIONS}

\noindent {\bf Fig. 1} The 3.4 cm continuum image of NGC 
3576 is shown at a resolution of 
7\arcsec~ (0.09 pc at 3.0 kpc).  
The first positive and negative continuum contours are at 3$\sigma$ (54 mJy
beam$^{-1}$).  Subsequent positive contours are 1.4, 2, 2.8, 4, 5.6, 8, 11.3, 16,
22.6, 32 and 45.3 times the 3$\sigma$ level.  The peak continuum flux
density is
4.1 Jy beam$^{-1}$
The positions of the 10 $\micron$ sources (Frogel \& Person 1974)
are indicated with crosses. Errors in the absolute IR positions are $\pm$4\arcsec (indicated by the cross size).
The inset is an integrated profile made over a
region of the source where broad recombination lines are detected. The
best fit to the spectrum consists of two Gaussians separated by $\sim$
25 \kms. Epoch is B1950.

\noindent {\bf Fig. 2} The 3.4 cm continuum image of NGC 3603 with the 13
named sources
(A to M in order of increasing RA).  The
continuum data has a spatial resolution of 7\arcsec~(0.21 pc at D = 6.1 $\pm$ 0.6
kpc).  The first
contour is at 5$\sigma$ (55 mJy beam$^{-1}$) with subsequent levels at
1.4, 2, 2.8, 4, 5.6, 8, 11.3, 16, 22.6, 32 and 45.3 times the 5$\sigma$
level.  The peak continuum flux density is 0.47 Jy beam$^{-1}$.
The positions of the 10 $\micron$ sources (Frogel et al.
1977) are indicated with crosses. Errors in the
absolute IR positions from this later study are $\pm$10\arcsec (indicated by the cross size). Epoch is B1950.

\noindent {\bf Fig. 3} (a) The continuum weighted, integrated line
profile for NGC 3576. The
recombination lines shown (left to right) are H113$\beta$, C90$\alpha$,
He90$\alpha$, and H90$\alpha$.
The solid line is the Gaussian fit, the
crosses are the data points, and the dashed line is the residual.  The
spectral resolution is 2.6 km s$^{-1}$.
(b) The continuum weighted, integrated line
profile for the bright southern region in NGC 3603. The
recombination lines shown (left to right) are H113$\beta$, C90$\alpha$,
He90$\alpha$, and H90$\alpha$.
The solid line is the Gaussian fit, the
crosses are the data points, and the dashed line is the residual.  The
spectral resolution is 2.6 km s$^{-1}$.

\noindent {\bf Fig. 4} The continuum weighted, integrated line profiles for
the Irs sources in NGC 3576 (Frogel \& Persson 1974), indicated in Figure 1.
The solid line is the Gaussian fit, the
crosses are the data points, and the dashed line is the residual.  The
spectral resolution is 2.6 km s$^{-1}$. Figures show (a) Irs 1/2,
(b) Irs 3, (c) Irs 4, (d) Irs 5.

\noindent {\bf Fig. 5} Line parameters of the H90$\alpha$ line in NGC 3576 overlaid on the
3.4 cm continuum contours. Fits were made to each pixel above the
5$\sigma$ level in the continuum.
(a) The velocity ($V_{LSR}$) of the H90$\alpha$ line in
NGC 3576 for each pixel is presented in grey-scale with the NGC 3576 3.4 cm
continuum contours overlayed.  The contour levels
are as indicated in Figure 1. The grey
scale velocity range is from -35 km s$^{-1}$ to 10 km s$^{-1}$.
(b) The full width at half maximum ($\Delta$$V_{FWHM}$) of 
the H90$\alpha$ line in
NGC 3576 for each pixel is presented in false color.
The line width range 
is from 20 km s$^{-1}$ to 50 km s$^{-1}$.
(c) The electron temperature calculated for each
pixel in NGC 3576 above 10$\sigma$ in the continuum
image.  The temperature (grey scale) 
ranges from 5000 to
12000 K.  The contours start at 4000 K and rise linearly to 12000 K
in increments of 1000 K.

\noindent 
{\bf Fig. 6} The integrated electron temperature in NGC 3576 for
18 concentric semicircles (width of 4 pixels) centered on the peak
in the 3.4 cm continuum image.  Temperatures range from 6000 K to
9000 K across the source.

\noindent
{\bf Fig. 7} The integrated line profiles from the 13 individual
HII regions in NGC 3603, as indicated in Figure 2.
The recombination lines shown (left to right where detected) 
are H113$\beta$, C90$\alpha$,
He90$\alpha$, and H90$\alpha$.
The plot shows the Gaussian fit (solid line), data (crosses),
and the residual (dashed line).
The spectral resolution is 2.6 km s$^{-1}$. Figures 7a to 7m 
show the integrated
profiles of sources NGC 3603 A-M respectively.

\noindent
{\bf Fig. 8} The 3.4 cm continuum contours for NGC 3576. The
stellar positions of cluster member stars with infrared excess 
(Persi et al. 1994) are plotted as crosses.
Also plotted are the positions of the star
HD 97499 (diamond), and the H$_2$O (triangle; Caswell 1989) and CH$_3$OH 
(box; Caswell et al. 1995) maser positions.
Continuum contours are as indicated in Figure 1. Epoch is J2000.

\noindent
{\bf Fig. 9} The 3.4 cm continuum contours in the vicinity of the
known stellar cluster HD 97950.
The positions of $\sim$50 cluster members with measured $U, B$, and $V$ 
magnitudes
(Melnick, Tapia \& Terlevich 1989) are plotted as crosses.
Coordinates of the 50 stars
were derived from pixel offsets from the known position of Sher 25
(a B1.5Iab supergiant)
as given by Brandner et al (1997). Also indicated is the position of
HD 97950 (diamond), and H$_2$O (triangles; Caswell 1989) and OH 
(box; Caswell 1998) maser positions.
Continuum contours are as indicated in Figure 2. Epoch is J2000.

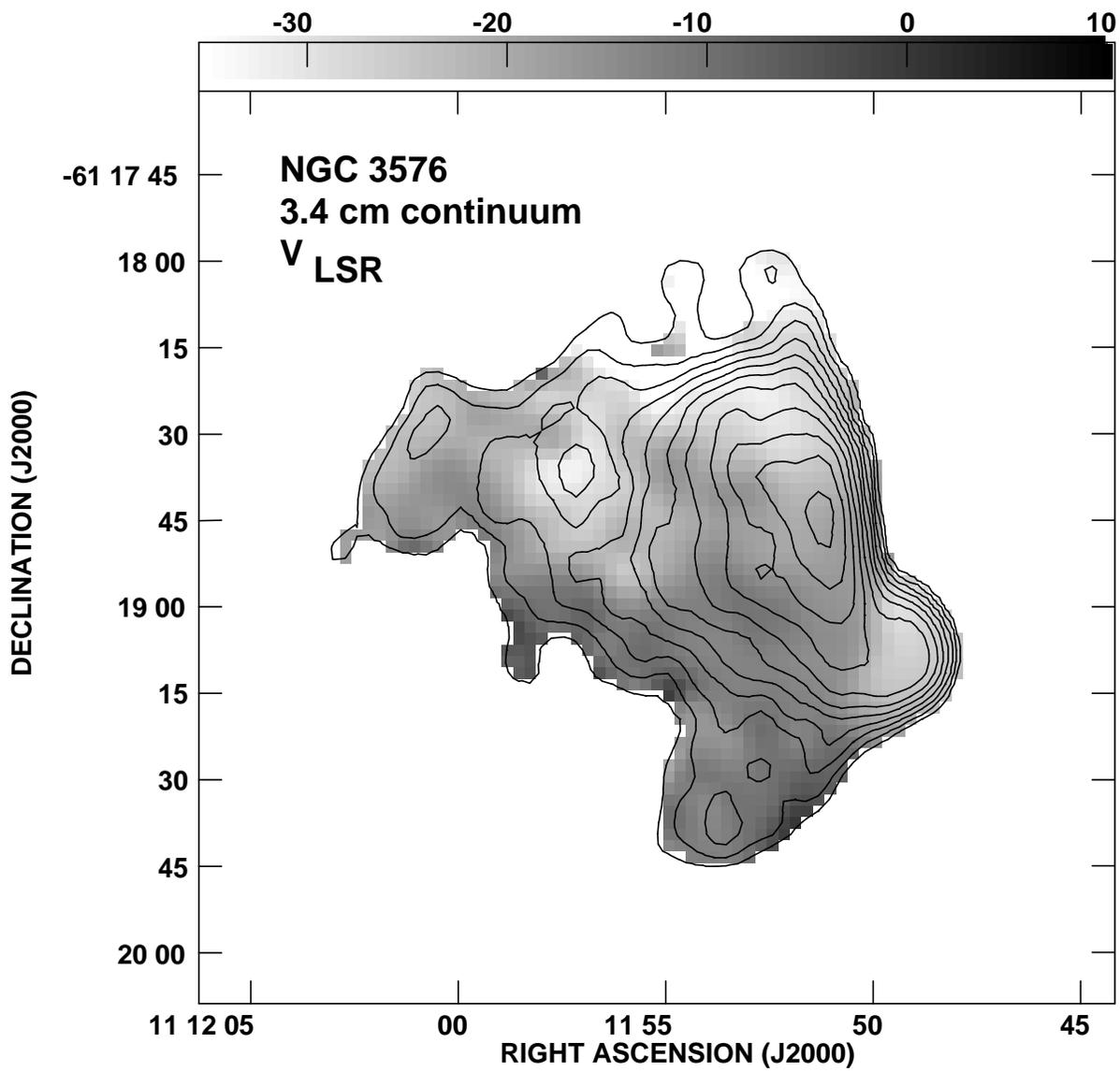
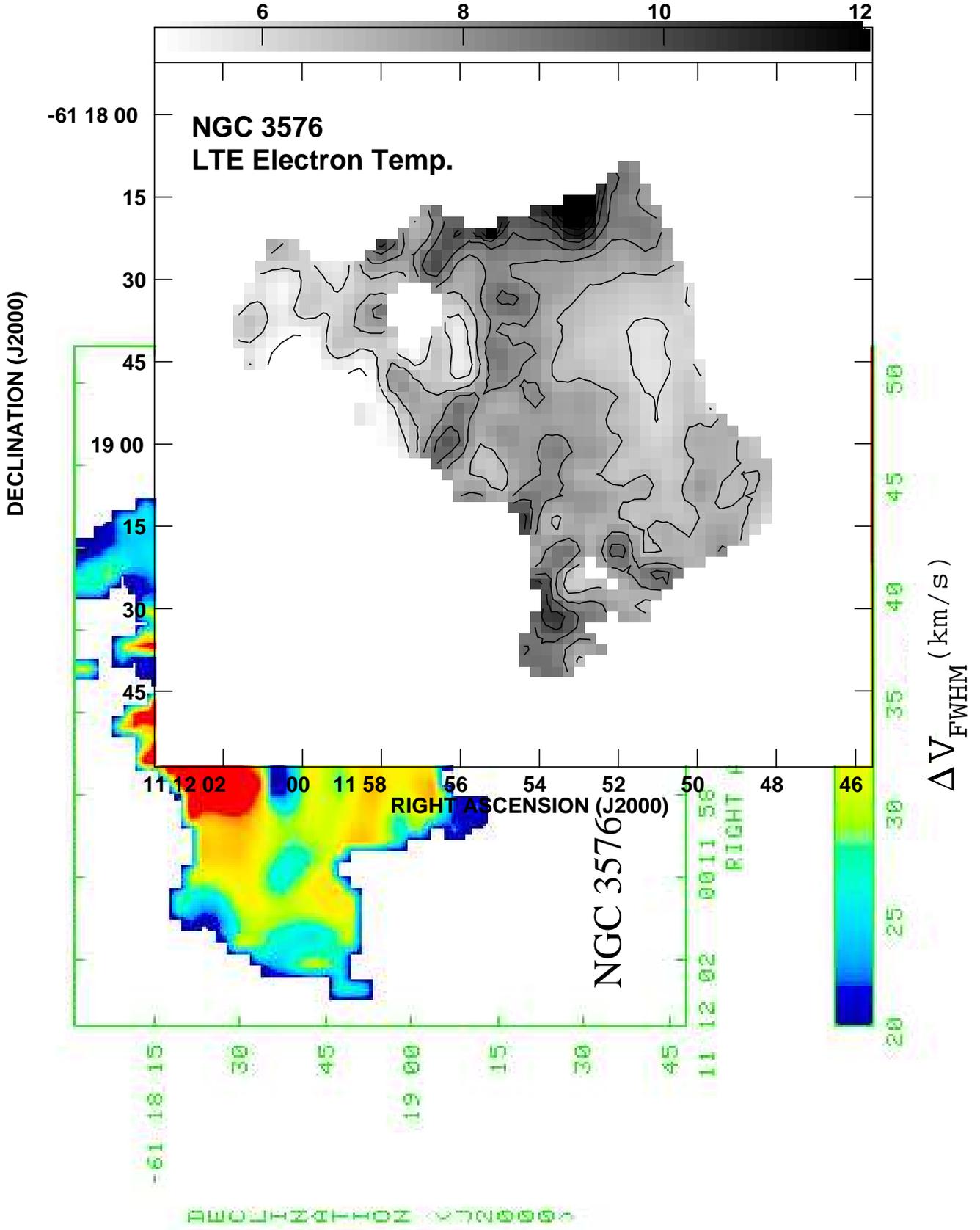
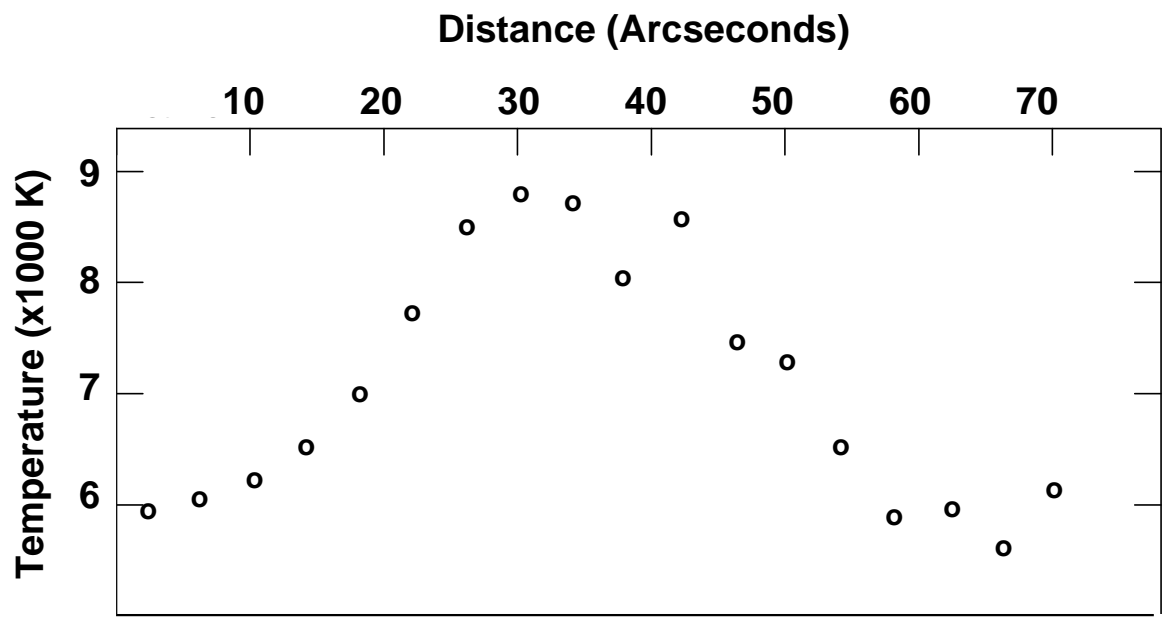
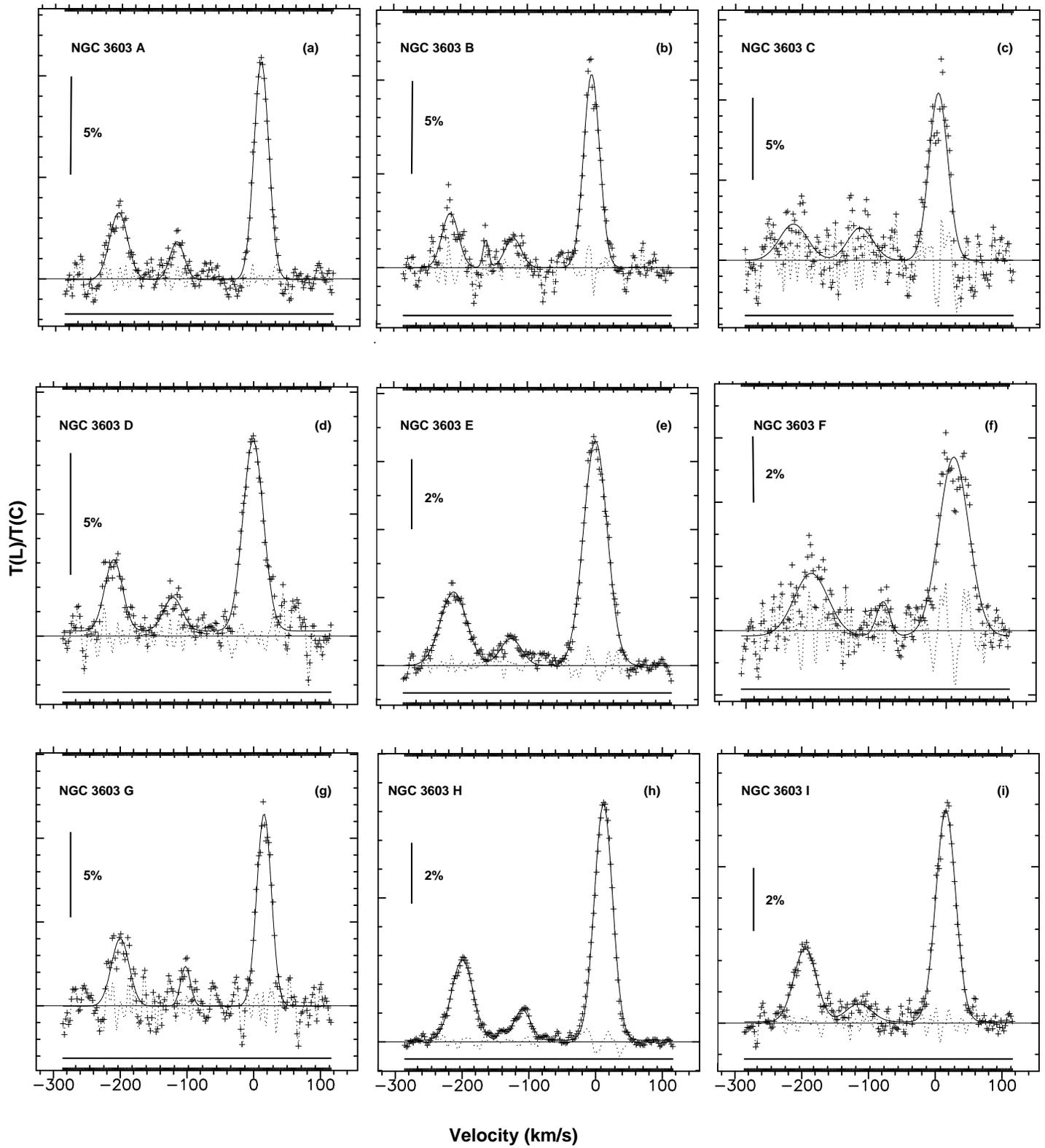
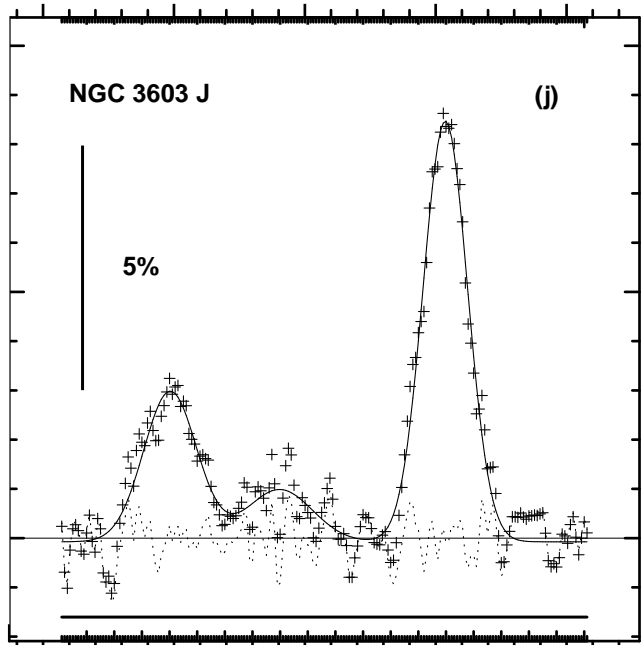
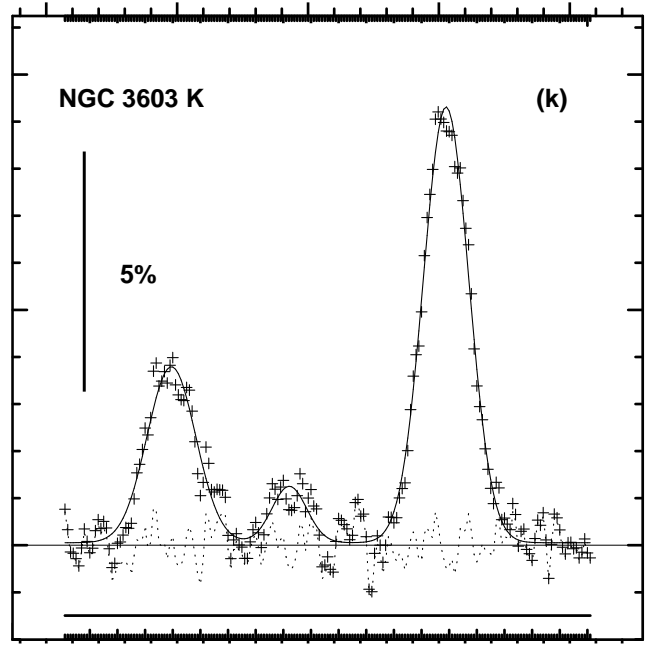
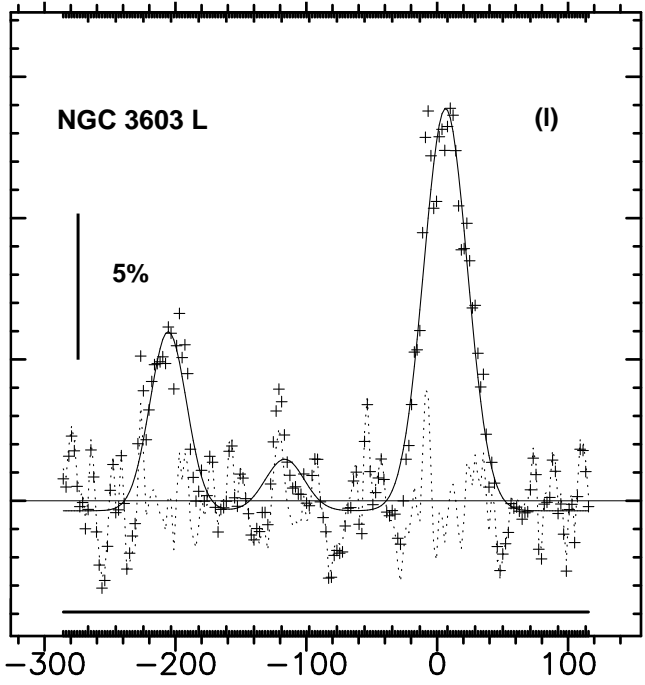
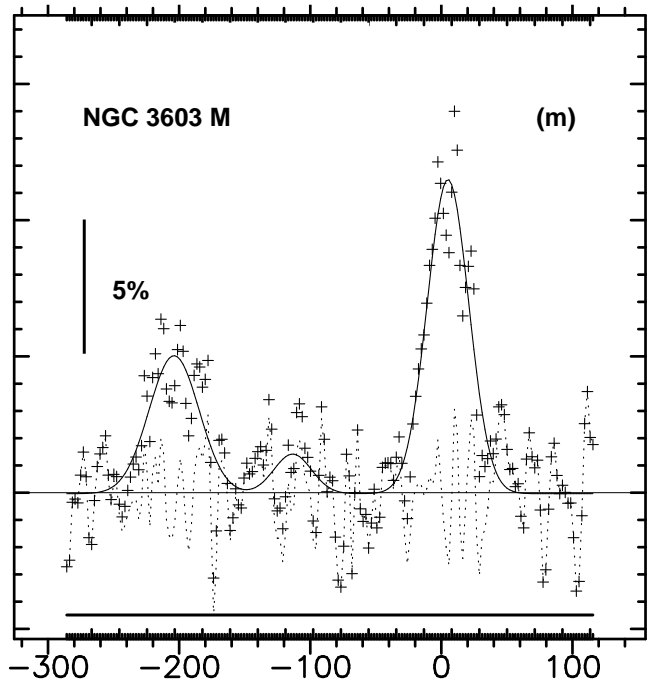
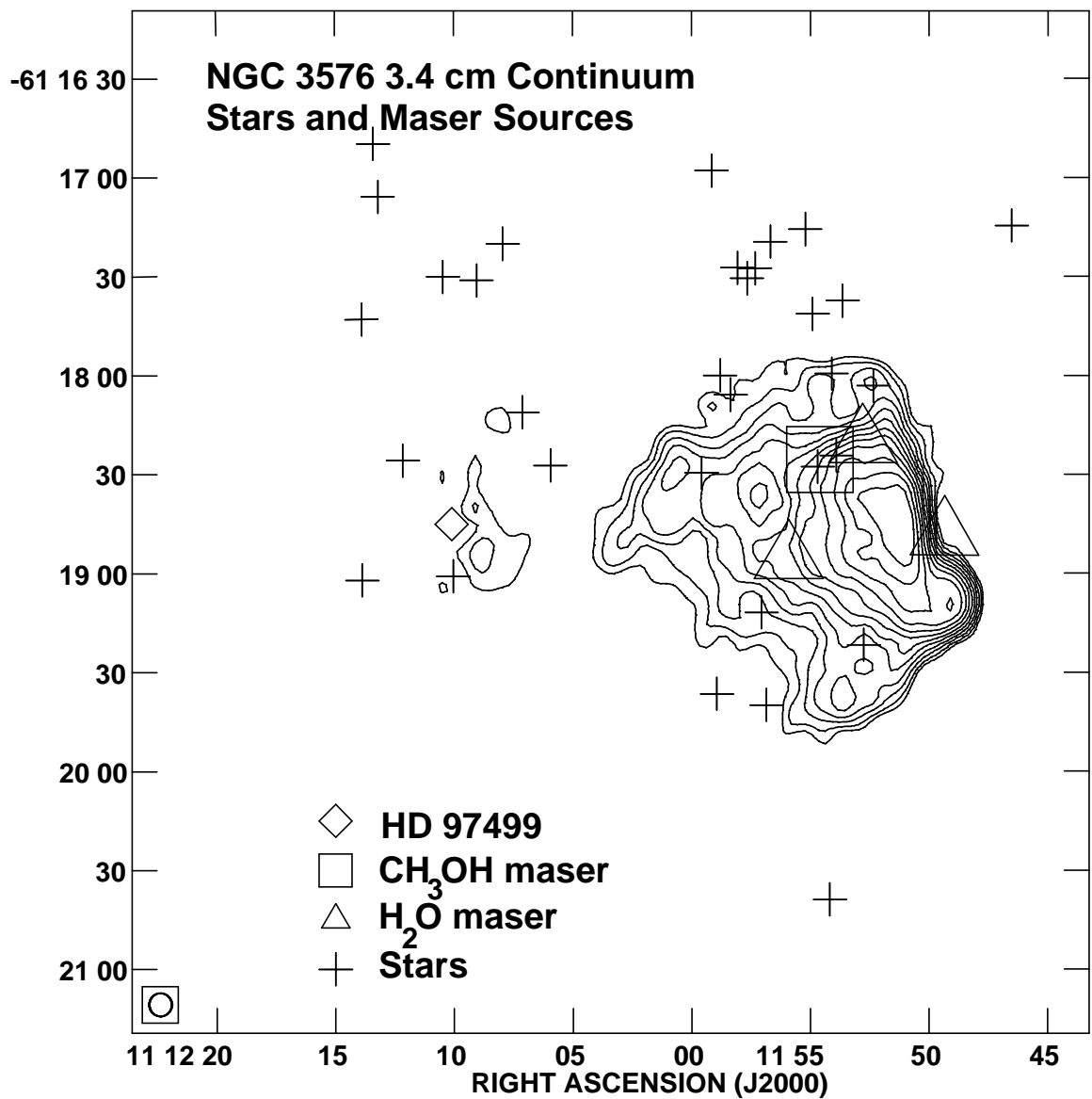
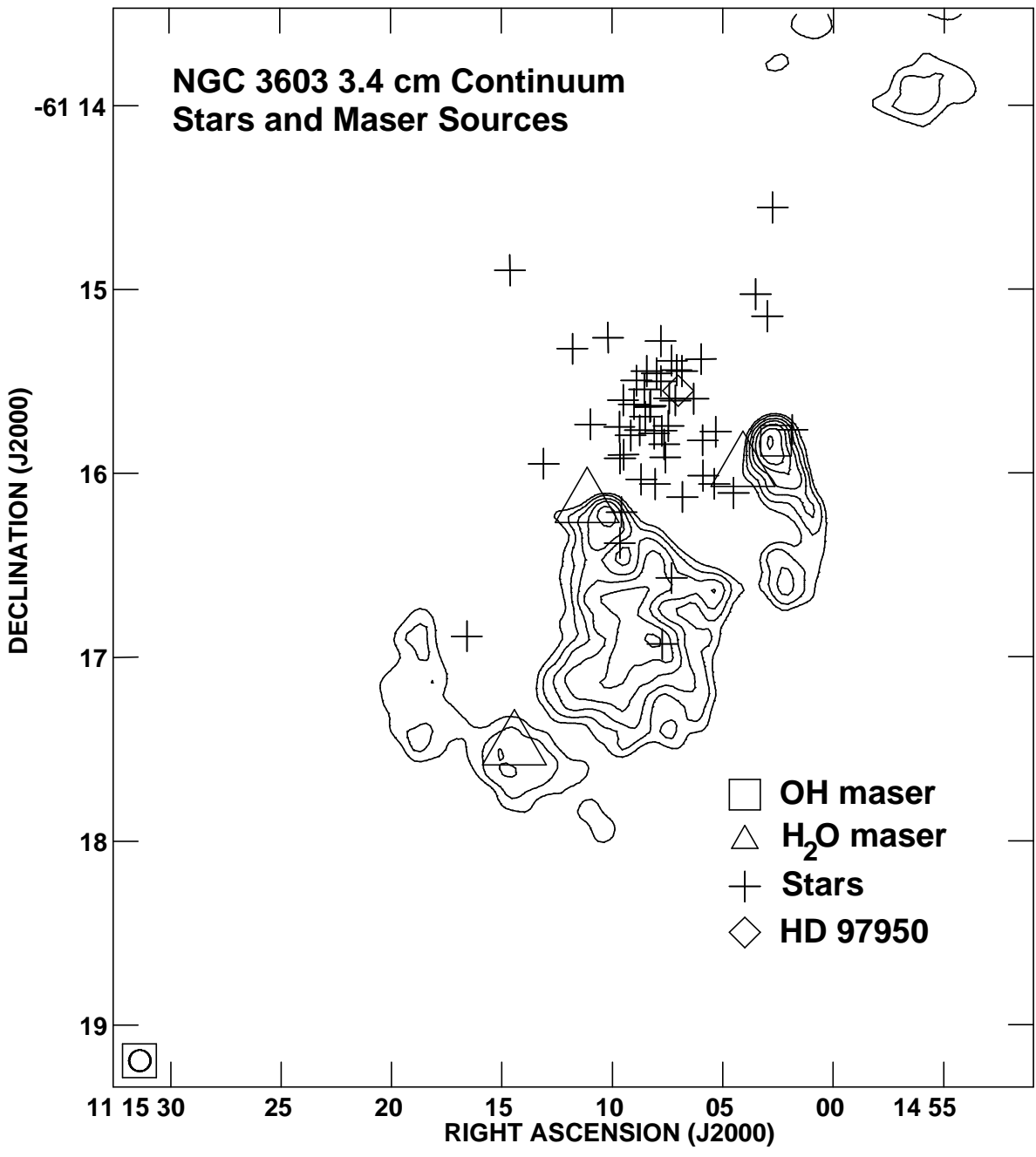
\end{document}